\documentclass[twocolumn,showpacs,preprintnumbers,pre]{revtex4}

\usepackage{amssymb}
\usepackage{amsmath}
\usepackage{graphicx}
\usepackage{color}
\usepackage{float}
\usepackage{subfigure}

\newcommand{\reffig}[1]{Fig.~\ref{#1}}
\newcommand{\refeq}[1]{Eq.~\ref{#1}}
\newcommand{\reftbl}[1]{Table~\ref{#1}}
\newcommand{\refsec}[1]{Sec.~\ref{#1}}

\begin{document}
\title{Gossip on Weighted Networks}
\author{Mursel Tasgin}
\author{Haluk O. Bingol}
\affiliation{Department of Computer Engineering\\
Bogazici University, Istanbul \\
}

\begin{abstract}
We investigate how suitable a weighted network is for gossip spreading. 
The proposed model is based on the gossip spreading model 
introduced by Lind et.al. on unweighted networks.
Weight represents ``friendship’’.
Potential spreader prefers not to spread if the victim of gossip is a ``close friend''.
Gossip spreading is related to the triangles and cascades of triangles.
It gives more insight about the structure of a network. 

We analyze gossip spreading on real weighted networks of human interactions. 
6 co-occurrence and 7 social pattern networks are investigated.
Gossip propagation is found to be  a good parameter to distinguish co-occurrence and social pattern networks. 
As a comparison some miscellaneous networks and 
computer generated networks based on ER, BA, WS models are also investigated.
They are found to be quite different than the human interaction networks.
\end{abstract}

\maketitle

\section{Introduction}

Gossip is one of the oldest and most common means of 
information sharing among people; 
in Greek mythology there is an icon named Pheme, 
who is a many-tongued character and initiates and furthers communication. 
Unlike rumor~\cite{Nekovee2007PA}, gossip is more personal and 
it is spread by the people who know the person being victimized by the gossip.
Most of the time the content of the speech is supposed to be secret 
among the ones 
who have heard the gossip however the one who gets this new information, 
whether he/she promised not to tell anyone else 
may further the gossip to other people. 
Eventually a secret or gossip becomes something widely known by 
many friends of the person who is the victim of the spread information.

Although gossip spreading concept has roots in social sciences 
other fields such as Computer Science has some interest.
In ad hoc networks, a routing algorithm, called gossip protocol, is inspired by gossip propagation in social systems~\cite{Demers1987}.

There have been recent studies about gossip spreading in complex 
networks~\cite{Lind2007EPL, Lind2007PRE}. 
The gossip spreading model proposed is based on 
information spreading among the first-degree neighborhood of the victim. 
The model is based on the assumption that 
gossip is personal and 
people tend to spread gossip about 
people they know 
to other people who also know the victim. 
When the model is applied to social networks, 
it is observed that there exists a degree $k_{0}$ such that
gossip spreading becomes minimum if the victim is of 
degree $k_{0}$~\cite{Lind2007EPL}.
Similar result are obtained for networks generated by Barabasi-Albert 
model~\cite{Barabasi1999Science}.
This paper extends the model to weighted networks.

\section{Gossip Spreading}
 
First we need same terminology.
A \emph{victim} $v$ is the node who is the subject of the gossip and 
will suffer from the spread of the gossip. 
A \emph{spreader} $s$ is the node 
who hears the gossip and furthers it. 
A \emph{target} $t$ is a node that is connected to both the victim and the spreader.
Then, gossip about the victim is spread in the network from spreader to target 
which in turn becomes a spreader.
The node which originates the gossip is called the \emph{originator} $r$.
The \emph{1-neighborhood} of node $i$, denoted by $N_{1}(i)$, is the set of nodes whose distance to $i$ is 1.
The degree of $i$ is $k_{i} = | N_{1}(i) |$.
Let $V$ and $E$ be the sets of nodes and edges, respectively.
$N = | V |$ and $M = | E |$.

A couple of observations is needed.
Note that victim-spreader-target forms a triangle.
How far the gossip would spread depends on the triangles in the topology.
First consider some extreme cases.
If the network is a complete graph,
all the nodes in $N_{1}(v)$ ``know'' each other and $v$,
hence any two nodes and $v$ makes a triangle.
Therefore all the nodes in $N_{1}(v)$ would get the gossip,
independent of who is the originator.
On the other hand, in a star connected network, 
no gossip about any node can spread since there is no triangle.

\begin{figure}
	\begin{center}
		\includegraphics[width=0.5\linewidth]
		{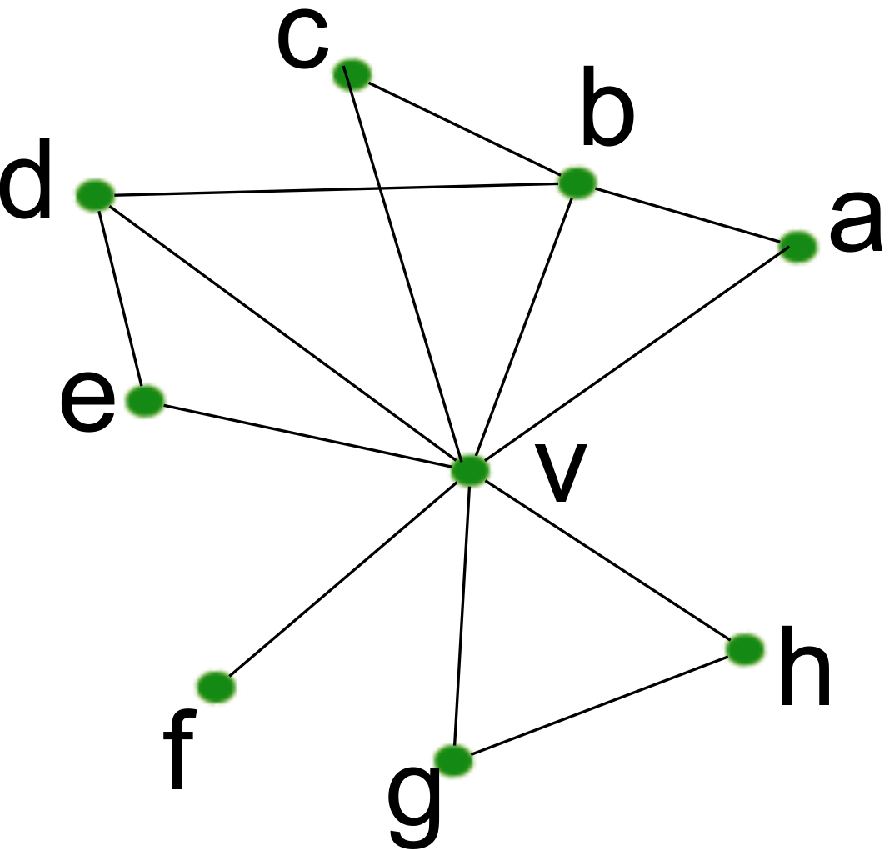}
		\caption{(Color online)
			A sample network for gossip propagation.
			For unweighted case
				$\sigma_{va} = \sigma_{vb}= \sigma_{ve} = 5/8$,
				$\sigma_{vf} = 1/8$,
				$\sigma_{vg} = 2/8$ and
				$\sigma_{v} = 30/64$.
	         For weighted case 
				$\beta_{vf} = \beta_{vb} = 1/8$,
				$\beta_{va} = \beta_{vg} = 2/8$,
				$\beta_{vd} = \beta_{ve} = 3/8$ and
				$\beta_{v} = 16/64$
				when all the weights $w_{ij} = 1$ except $w_{vb} = 2$.  
		}
		\label{fig:SampleNetworkGossip}
	\end{center}
\end{figure}

In a typical network some 1-neighbors are connected to each other.
The connected 1-neighbor can gossip
yet the degree of spread depends on the selection of the originator.
A simple network is given in \reffig{fig:SampleNetworkGossip} .
Let $v$ be the victim.
No gossip occurs when $f$ is the originator. 
If $g$ is the originator, only $h$ can get the gossip.
Actually due to $v-g-h$ triangle, there are two possible gossips: one from $g$ to $h$ and the other from $h$ to $g$.
Consider the remaining nodes, $\{a, b, c, d, e\}$.
If any one is selected as the originator, 
eventually the rest gets the gossip.
Suppose $b$ is the originator.
$a, c,d$ get it immediately since 
each makes a triangle with the common side $v-b$,
i.e. $a, c, d \in N_{1}(v) \cup N_{1}(b)$.
Although $a, c$ cannot propagate it any further, $d$ can.
Once $d$ gets it, it propagates to $e$.
Here a \emph{gossip cascade} occurs since 
the common side changes to $v-d$.

As a summary 
(i) a triangle is require for a single gossip, 
(ii) a sequence of triangles with pairwise common edge is necessary for gossip cascades.

Note that given a network, 
the size and the duration of the gossip propagation depends on 
who is the victim and who originates it, $v$ and $r$, respectively. 
After the propagation finishes, two metrics are investigated~\cite{Lind2007EPL}.
\emph{Spreading time} $\tau_{vr}$ is 
the largest distance between $r$ and nodes that receive the gossip about $v$.
\emph{Spread-factor} $\sigma_{vr}$ is the fraction the friends $n_{vr}$ of $v$
who received the gossip, that is,
\[
	\sigma_{vr} = \frac{n_{vr}}{k_{v}}.
\]
Then one can define the following averages 
\begin{eqnarray}
	\sigma_{v} &= &\frac{1}{k_{v}} \sum_{r \in N_{1}(v)} \sigma_{vr}, \\
	\sigma &= &\frac{1}{N} \sum_{v \in V} \sigma_{v} \label{eq:sigma}.
\end{eqnarray}
In order to observe the existence of $k_{0}$, 
vertices of the same degree are considered:
\[
	\sigma_{k_{i}} = \frac{1}{|V_{k_{i}}|} \sum_{v \in V_{k_{i}}} \sigma_{v}
\]
where $V_{k_{i}}$ is the set of vertices with degree $k_{i}$.
Note that $\sigma_{vr}, \sigma_{v}, \sigma \in [0, 1]$ 
since $0 \leq n_{vr} \leq k_{v}$.

\section{The Model}

The base model in ref~\cite{Lind2007EPL} is defined on unweighted networks. 
A node must propagate gossip once it gets it. 
In this model, node does not have any decision on propagation,
therefore gossip can spread as far as the network topology permits.
So the connectivity decides.
In ref~\cite{Lind2007PRE} the model is extended to the case 
where spreader decides to spread with probability $p$.

In this work we assume that
the receiver of gossip 
neither blindly propagates it 
nor use a probability in propagation. 
Spread or stop decision is usually based on 
how closely related the spreader and the victim~\cite{Shaw2010Complexity}.
If the victim is a close friend of the spreader,
he prefers to not propagate the gossip.
With this motivation 
we propose a gossip spreading model that extends that of 
ref~\cite{Lind2007EPL} in two  ways:
(i)~The underlining network is weighted network where 
weight $w_{ij}$ represents the strength of the ``friendship'' 
between nodes $i$ and $j$.
(ii)~Spreader decides whether stop or propagate the gossip 
based on its ``friendship'' with the victim.
The spreading is based on triangle cascades as in ref~\cite{Lind2007EPL} running on the corresponding unweighted network.
The difference is that spreader may choose to stop propagation 
if he thinks that victim is a close friend.
We define close friendship as being closer than the average, that is,
node $v$ is a \emph{close friend} of node $s$ if
\[
	w_{sv} > \frac{1}{k_{s}} \sum_{\ell \in N_{1}(s)} w_{s \ell}.
\]

Note that decision of $s$ to propagate a gossip about $v$ depends on not only their friendship $w_{sv}$ but also 1-neighborhood of $s$.
If $s$ has closer friends then $v$ it will gossip about $v$.


In order to quantify the spread of gossip
we extend $\sigma$-metric to corresponding $\beta$-metric as follows:
\begin{eqnarray}
	\beta_{vr} &= &\frac{m_{vr}}{k_{v}}, \\
	\beta_{v} &= &\frac{1}{k_{v}} \sum_{r \in N_{1}(v)} \beta_{vr}, \\
	\beta &= &\frac{1}{N} \sum_{v \in V} \beta_{v} 	\label{eq:beta}
\end{eqnarray}
where $m_{vr}$ is the number of friends of $v$ 
who receives the gossip originated by $r$ in the weighted network.
Note that considering a weighed network and its corresponding unweighted network, $0 \leq m_{vr} \leq n_{vr}$ 
since some nodes in $N_{1}(v)$ in the weighted network prefer not to propagate 
while the corresponding nodes in the unweighted network always propagate.
Therefore $\beta_{vr}, \beta_{v}, \beta \in [0, 1]$.

For $V_{k_{i}}$ being the set of vertices with degree $k_{i}$
\[
	\beta_{k_{i}} = \frac{1}{|V_{k_{i}}|} \sum_{v \in V_{k_{i}}} \beta_{v}.
\]

\begin{figure}
\begin{center}
	\includegraphics[width=0.5\linewidth]
	{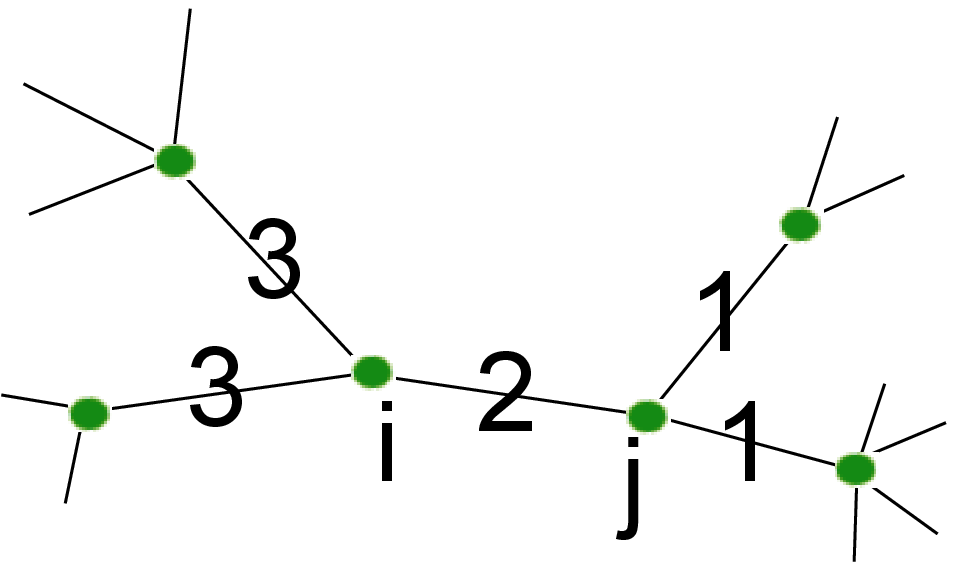}
	\caption{(Color online)
		Gossip is not symmetric.
		The average friendship of $a$ and $b$ are 
		$11/2$ and $4/3$, respectively.
		Since $w_{ab}=2$, $i$ can propagate gossip about $j$ 
		while $j$ does not.                    
	}
	\label{fig:SampleNetworkGossipNotSymmetric}
\end{center}
\end{figure}
Note that to close friend relation is not symmetric.
Consider a network segment is given in 
\reffig{fig:SampleNetworkGossipNotSymmetric}.
Node $i$ can gossip about node $j$ 
since the average friendship of $i$ is $11/2$ and friendship of $i-j$ is only 2.
On the other hand since $j$ does not have good ties, its average friendship is $4/3$. Therefore $j$ considers $i$ as a close friend and prefers not to gossip about it.
In this respect the proposed model is actually defined on directed weighted networks.
Because of that every edge has to be considered twice for one end gossiping about the other.


\section{Data Sets}
The data sets used in ref~\cite{Lind2007EPL, Lind2007PRE} cannot be used since 
they is unweighted.
The unweighted base model of ref~\cite{Lind2007EPL} and 
the proposed model are applied on a number of undirected weighted networks.
Although the proposed model also applies to directed weighted networks,
we leave directed weighted networks for another study.
Since gossip is the focus of this work, 
data sets related to human are selected. 
They have come from two basic characteristics, namely, co-occurrence networks and social pattern networks.
In order to compare, data sets of different characters: one from linguistics, the other one from neuroscience are also used.

\textbf{Co-occurrence networks.}
Co-occurrence networks are based on bipartite graphs $G(A \cup B, E)$ 
where the set of nodes are in dichotomy of $A, B$. 
The nodes of the corresponding co-occurrence graph are vertices in $A$.
Nodes $v_{i}, v_{j} \in A$ are connected whenever 
there is $b \in B$ such that $v_{i}$ and $v_{j}$ are connected to $b$ in the bipartite graph $G$.

Reuters-21578 corpus is well-know in Computer Science literature.
It composed of 21,578 Reuters news articles in 1987.
In Reuters co-occurrence network, denoted by CREU, nodes are person appear in 
the articles~\cite{Ozgur2004LNCS}. 

Two person are connected if they appear in the same article.
Weight $w_{ij}$ is defined as the number of times two person appeared 
in the same article together.
Note that each article contributes $n$ to the sum of weights, i.e. $\frac{1}{2} \sum_{i,j} w_{ij}$, where $n$ is the number of persons occur in the article.

In co-authorship networks authors are represented by nodes.
Two authors are connected if they have a common paper.
Every paper with $n$ author contributes $1/(n-1)$ to 
the weight associated to an edge between two of its authors.
Note that each paper has a total contribution of $n/(n-1)$
to the sum of weights.
We investigate the co-authorship networks of 
High-Energy Physics Theory, 
Condensed Matter collaborations 2005, 
Astrophysics~\cite{Newman2001PNAS} and 
coauthorship in Network Science~\cite{Newman2006PREeigenvectors},
denoted by
CPHE, 
CPCM, 
CPA, 
CNS, 
respectively.

Our final co-occurrence data set comes from quite a different 
domain~\cite{Knuth1993GraphBase}. The network is based on 
Victor Hugo's novel Les Miserables and the nodes represent key characters of the it.
Two nodes have a connection with each other if they co-appear on the same stage and weights on edges represent frequencies of their co-appearance.

\textbf{Social Pattern Networks.}
The SocioPatterns project (http://www.sociopatterns.org) collects data on socially interacting people in different settings. 
Nodes are the individuals.
Two individuals are connected by an weighted edge if 
they happen to be in ``closed-range face-to-face proximity''.	
Each edge has a weight which gives the duration of contact as the number of 20-second intervals.

The data sets, that are used in this work, are collected 
in a school environment,
in a scientific conference and 
in a long-running museum exhibition.
There are many days of recordings in the museum set.
We used data recorded on dates 
April 28, 
May 03, 
Jun 04 and 
July 07, 2009, 
denoted by 
SM0428, 
SM0503, 
SM0604, and 
SM0707, 
respectively~\cite{Isella2010}. 
The conference data is represented by 
SCON~\cite{Isella2010}.
There are two days of dataset for school environment. 
We denote the dataset of first day as 
SCH01 and second as 
SCH02~\cite{Stehle2011}.

\textbf{Miscellaneous Networks.}
In order to compare co-occurrence and social pattern networks, 
we consider two weighted networks from very different domains, namely, 
linguistics and neuroscience.
In simple terms given a word and asking subject to provide the first word that
comes to her mind is how word association is collected.
In a multi subject test, the frequency of association between two words is 
the weight of the edge connecting the two.
We use the data known as the Edinburgh Associative Thesaurus,
denoted by 
EAT,
which is an example of association network ~\cite{Kiss1973ESR}.

Neural network, 
denoted by
NCE, 
of the nervous system of C. Elegans has 302 neurons where neurons connects to neurons~\cite{Watts1998Nature, White1986RSB}.

\textbf{Genereted Networks.}
Finally, networks generated by means of well-known models of 
Erdos-Renyi,
Barabasi-Albert and
Watts-Strogatz are used~\cite{Erdos1959RandomGraphs,Barabasi1999Science, Watts1998Nature}.
Since the models generate unweighted networks, 
edge weights are assigned using a distribution.
The details are discussed in \refsec{sec:GeneratedWeightedNetworks}.

\section{Discussion}

%
%
\begin{table*}
	\caption{
		Network Coefficients\\
		($N$: number of nodes,
		$M$: number of edges,
		$k_{0}$: critical degree,
		$k_{0}^{w}$ : critical degree in weighted networks,
		$CC$: clustering coefficient,
		$\sigma$: spread factor in unweighted networks,
		$\beta$: spread factor in weighted networks.)
	}
	\begin{center}
	\begin{tabular}{|l|l|rr|rr|r|rrr|rrrr|r|}

\hline Type &Data Set&N&M&$k_{0}$&$k_{0}^{w}$&$\frac{k_{0}^{w}}{k_{0}}$&CC&$\sigma$&$\beta$&$\frac{\sigma}{\text{CC}}$&$\frac{\beta}{\text{CC}}$&$\frac{\beta}{\sigma}$&$\frac{\beta}{\sigma \text{ CC}}$&Ref\\

\hline \hline Co-occurrence&CNS&1,589&2,742&12&34&2.83&0.64&0.68&0.35&1.06&0.55&0.51&0.80&\cite{Newman2006PREeigenvectors}\\
Co-occurrence&CPHE&8,361&15,751&15&34&2.27&0.44&0.55&0.37&1.25&0.84&0.67&1.52&\cite{Newman2001PNAS}\\
Co-occurrence&CPA&16,706&121,251&11&37&3.36&0.64&0.79&0.48&1.23&0.75&0.61&0.95&\cite{Newman2001PNAS}\\
Co-occurrence&CPCM&40,421&175,691&27&69&2.56&0.64&0.78&0.49&1.22&0.77&0.63&0.98&\cite{Newman2001PNAS}\\
Co-occurrence&CLM&77&254&4&15&3.75&0.57&0.72&0.48&1.26&0.84&0.67&1.18&\cite{Knuth1993GraphBase}\\
Co-occurrence&CREU&5,249&7,528&21&34&1.62&0.44&0.47&0.24&1.07&0.55&0.51&1.16&\cite{Ozgur2004LNCS}\\
\hline \hline Social Pattern&SCON&113&2,196&NA&30&NA&0.53&0.99&0.82&1.87&1.55&0.83&1.57&\cite{Isella2010}\\
Social Pattern&SM0604&133&580&7&6&0.86&0.50&0.72&0.51&1.44&1.02&0.71&1.42&\cite{Isella2010}\\
Social Pattern&SM0428&206&714&4&6&1.50&0.41&0.71&0.49&1.73&1.20&0.69&1.68&\cite{Isella2010}\\
Social Pattern&SM0503&309&1,924&3&7&2.33&0.36&0.86&0.61&2.39&1.69&0.71&1.97&\cite{Isella2010}\\
Social Pattern&SM0715&422&2,841&5&5&1.00&0.45&0.82&0.52&1.82&1.16&0.63&1.40&\cite{Isella2010}\\
Social Pattern&SCH01&236&5,899&32&36&1.13&0.50&1.00&0.77&2.00&1.54&0.77&1.54&\cite{Stehle2011}\\
Social Pattern&SCH02&238&5,539&21&36&1.71&0.56&1.00&0.74&1.79&1.32&0.74&1.32&\cite{Stehle2011}\\
\hline \hline Miscellaneous&MNCE&297&2,148&5&32&6.40&0.29&0.81&0.50&2.79&1.72&0.62&2.14&\cite{Watts1998Nature}\\
Miscellaneous&MEAT&23,219&304,934&14&17&1.21&0.10&0.48&0.37&4.80&3.70&0.77&7.70&\cite{Kiss1973ESR}\\
\hline \hline Generated&GER&1000&10492&30&34&1.13&0.02&0.08&0.07&4.00&3.50&0.88&44.00&\cite{Erdos1959RandomGraphs}\\
Generated&GBA&1000&10380&37&46&1.24&0.13&0.42&0.26&3.23&2.00&0.62&4.77&\cite{Barabasi1999Science}\\
Generated&GWS&1000&10000&27&24&0.89&0.54&0.85&0.50&1.57&0.93&0.59&1.09&\cite{Watts1998Nature}\\
\hline

	\end{tabular}
	\end{center}
	\label{tbl:coefficients}
\end{table*}

We use weighted networks for our model.
Then the underlining unweighted network is used for the model of 
ref~\cite{Lind2007EPL}.
Comparison of our model with that of ref~\cite{Lind2007EPL} produces similarities and differences as summarized in \reftbl{tbl:coefficients}.

Data is organized in three groups in \reftbl{tbl:coefficients}.
The first group with 6 data sets are co-occurrence networks.
The second group of 7 data sets are social pattern networks.
The third group has data sets from different characters: one from linguistics, the other one from neuroscience.
The last  group is generated networks.

In \reftbl{tbl:coefficients}, the first two columns are the data sets.
$N$ and $M$ columns are the number of nodes and edges, respectively.
CC is the clustering coefficient~\cite{Watts1998Nature}.
$\sigma$ and $\beta$ are defined in \refeq{eq:sigma} and \refeq{eq:beta}, respectively.
$k_{0}$ and $k_{0}^{w}$ are the degrees where the gossip spread becomes minimum.

\subsection{Degree with Minimum Gossip Spread}
\begin{figure}
	\begin{center}
	\subfigure[CPCM]{
		\includegraphics[scale=0.4]{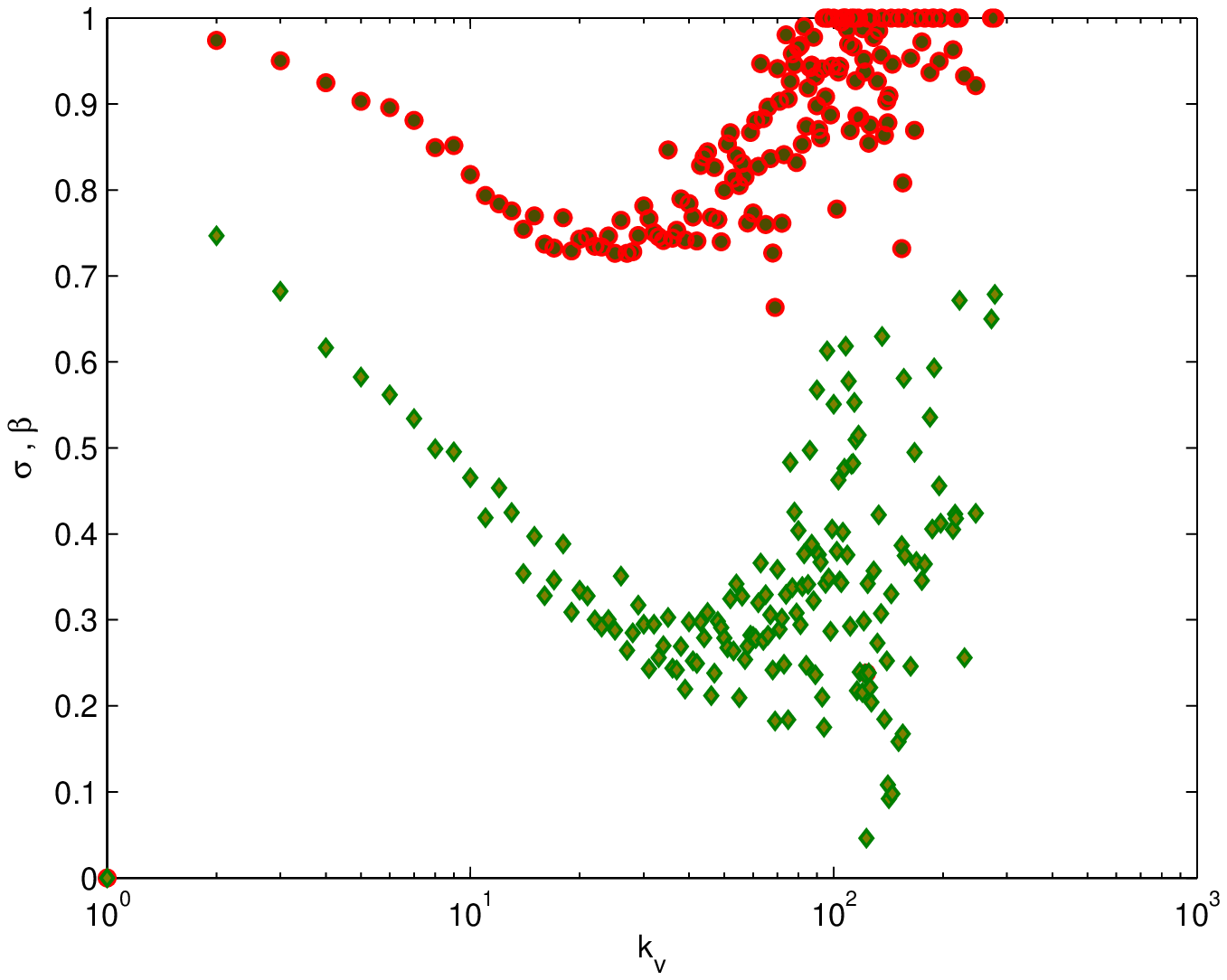}}
	\subfigure[CREU]{
	\includegraphics[scale=0.4]{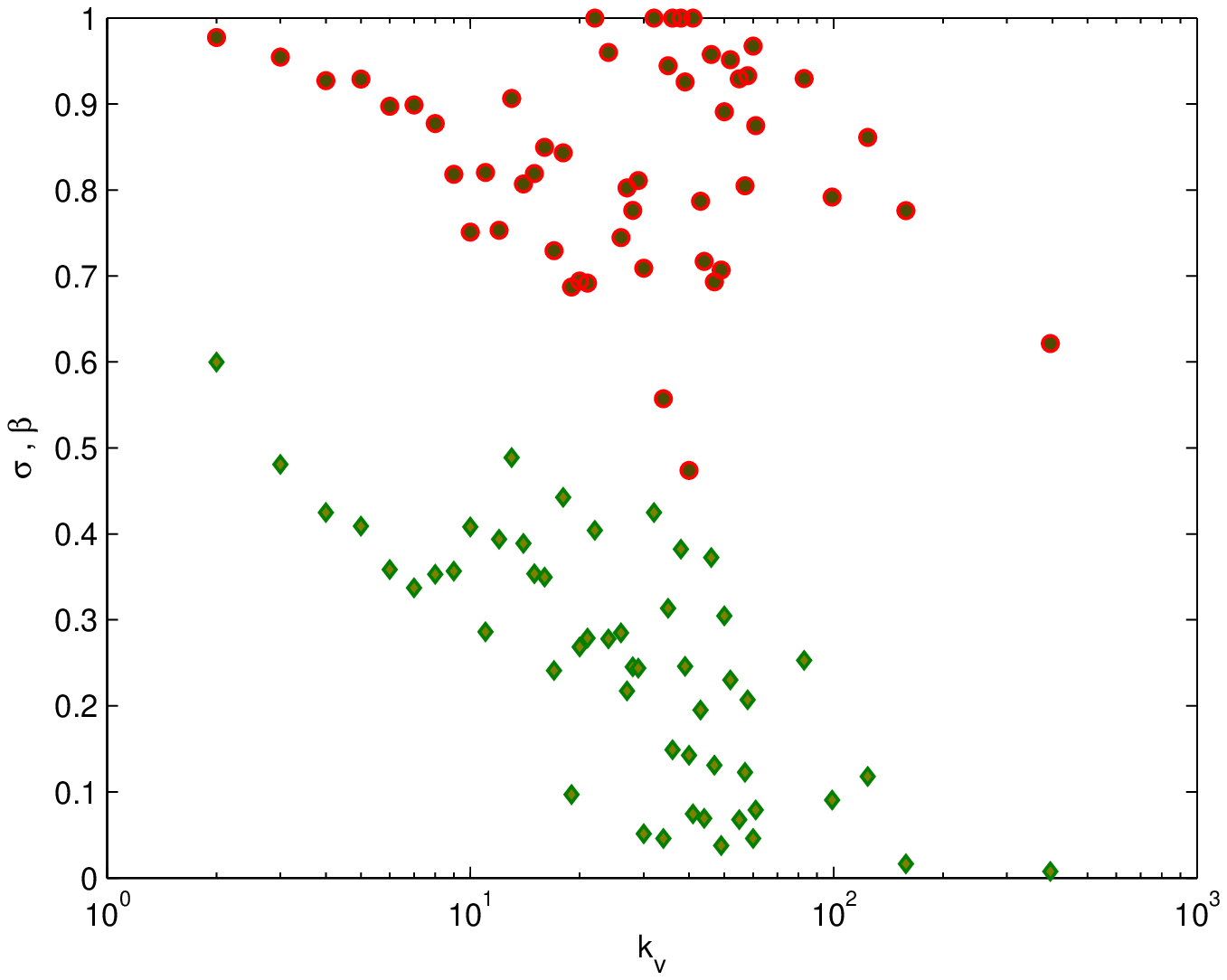}}
	\subfigure[SCON]{
	\includegraphics[scale=0.4]{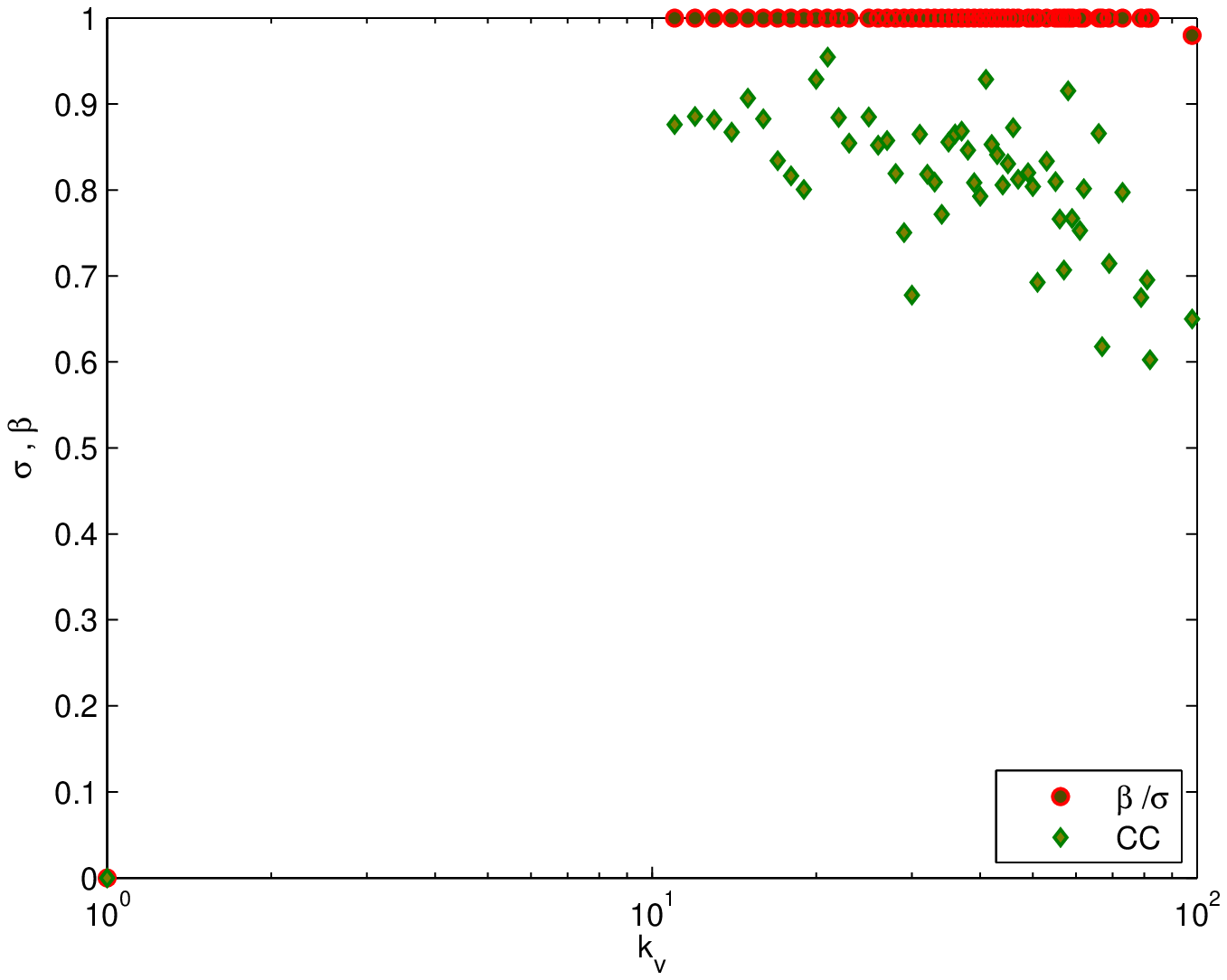}}

	\caption{(Color online)
		Gossip spread factors $\sigma_{k}$ and $\beta_{k}$ 
		as a function of $k$
		where $k$ is the degree of the victim.
	}
	\label{fig:k0}
	\end{center}
\end{figure}
One of the unexpected findings in ref~\cite{Lind2007EPL} is 
the existence of a degree $k_{0}$ where 
gossip spreading gets to a minimum.

First we want to check this observations on the networks that we work with.
Our model is defined on weighted networks.
The corresponding unweighted networks are obtained by 
removing the weights while keeping the connectivity.
When the corresponding unweighted networks are investigated,
such a minimum is observed in many of them.
\reffig{fig:k0} provides $\sigma_{k_{v}}$ and $\beta_{k_{v}}$ values as a function of degree of the victim $k_{v}$.
In the undirected network, the spread factor $\sigma_{k_{v}}$ starts as high, then decreases as $k_{v}$ increases from $0$ to some critical value $k_{0}$. 
As $k_{v}$ further increases $\sigma_{k_{v}}$ starts to increase 
as seen in networks CPCM, CREU  in \reffig{fig:k0}.
Note that social pattern network SCON is an exception.

The spread factor $\beta_{k_{v}}$  of a weighted network is always smaller 
in value than 
the spread factor $\sigma_{k_{v}}$ of the corresponding unweighted network 
since some gossip propagation that is possible in the unweighted network are blocked in the weighted counterpart due to spreader feels that victim is a ``close friend'' and stops spreading.
This can be observed in all graphs of \reffig{fig:k0}. 

When the weighted networks are considered and 
the spread factor $\beta_{k_{v}}$ of weighted networks are evaluated, 
a similar pattern is observed, that is,
we observe a degree $k_{0}^{w}$ where gossip spread is a minimum.
Although unweighted counter part does not have $k_{0}$,
we can see a subtle $k_{0}^{w}$ in SCON network for the weighted case.
$k_{0}^{w}$ values in CREU and SCON networks follows similar patterns for higher degrees.
This is due to the reductionist effect of weight distribution of highly connected nodes in these networks (i.e. having strong connections with other nodes such that they decide not to spread a gossip about a highly connected node).

It is also observed that this degree is always higher in the weighted networks,
i.e. $k_{0} < k_{0}^{w}$ as seen in \reftbl{tbl:coefficients}.
The ration $k_{0}^{w} / k_{0}$ gives an indication of the type of network.
For co-occurrence networks, the ratio is high. 
Except Reuters network CREU it is above 2. 
On the contrary, for social pattern networks it is low, around 1.
SM0503 network with 2.33 is an exception.

%
\subsection{Discriminating Co-occurrence and Social Pattern Networks}
%
%
\begin{figure}
	\begin{center}
	\subfigure[CPCM]{
	\includegraphics[scale=0.4]{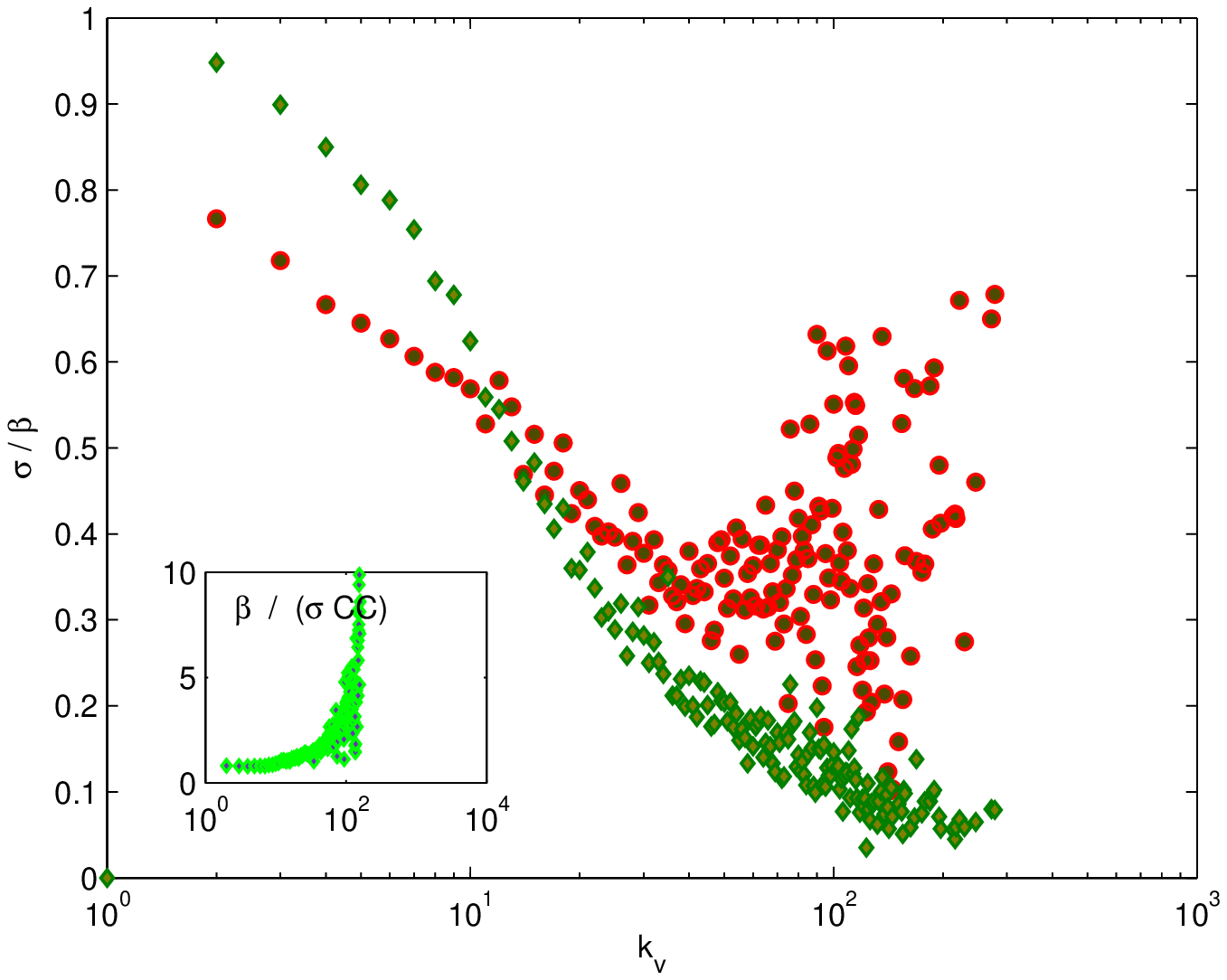}}
	\subfigure[CREU]{
	\includegraphics[scale=0.4]{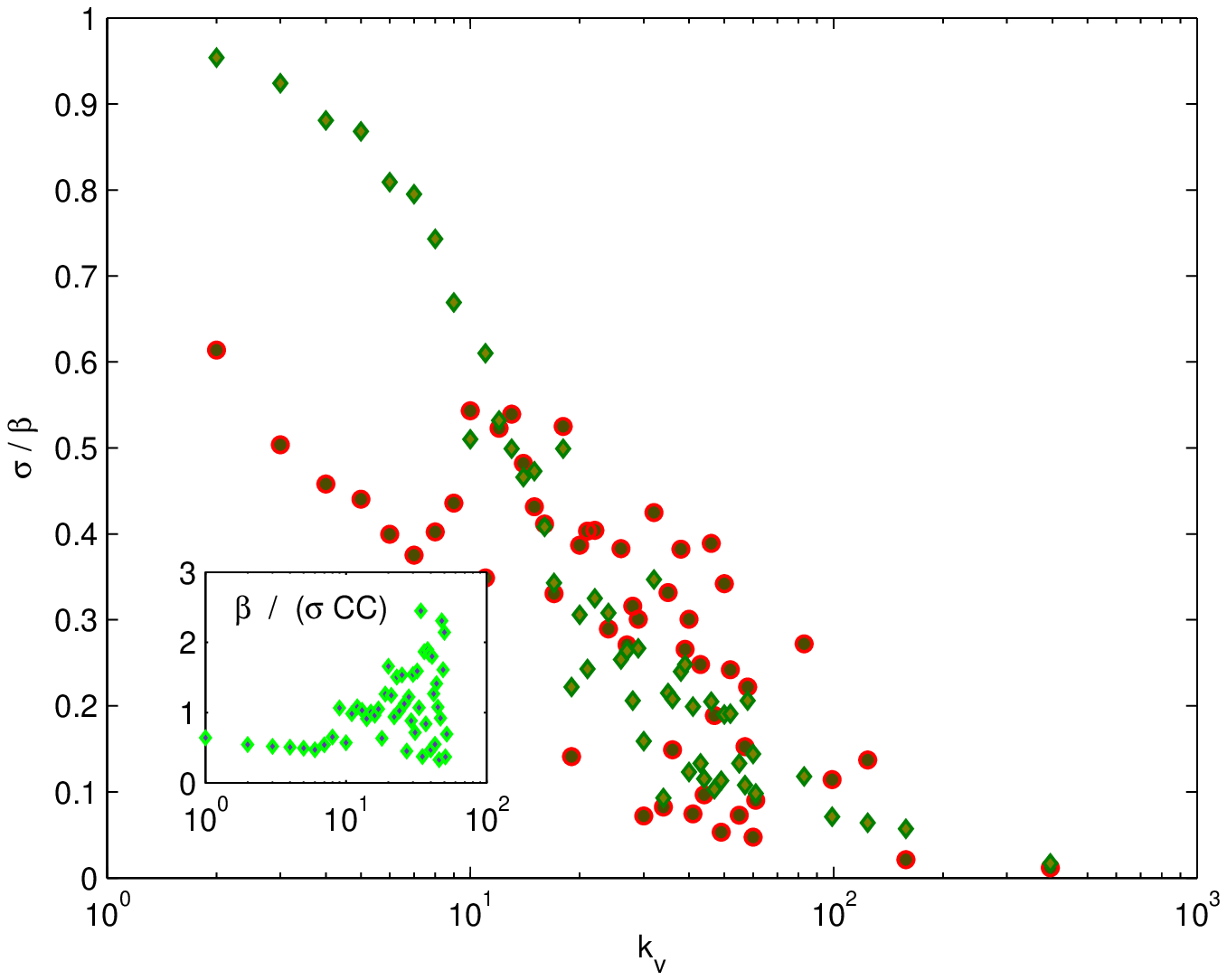}}
	\subfigure[SCON]{
	\includegraphics[scale=0.4]{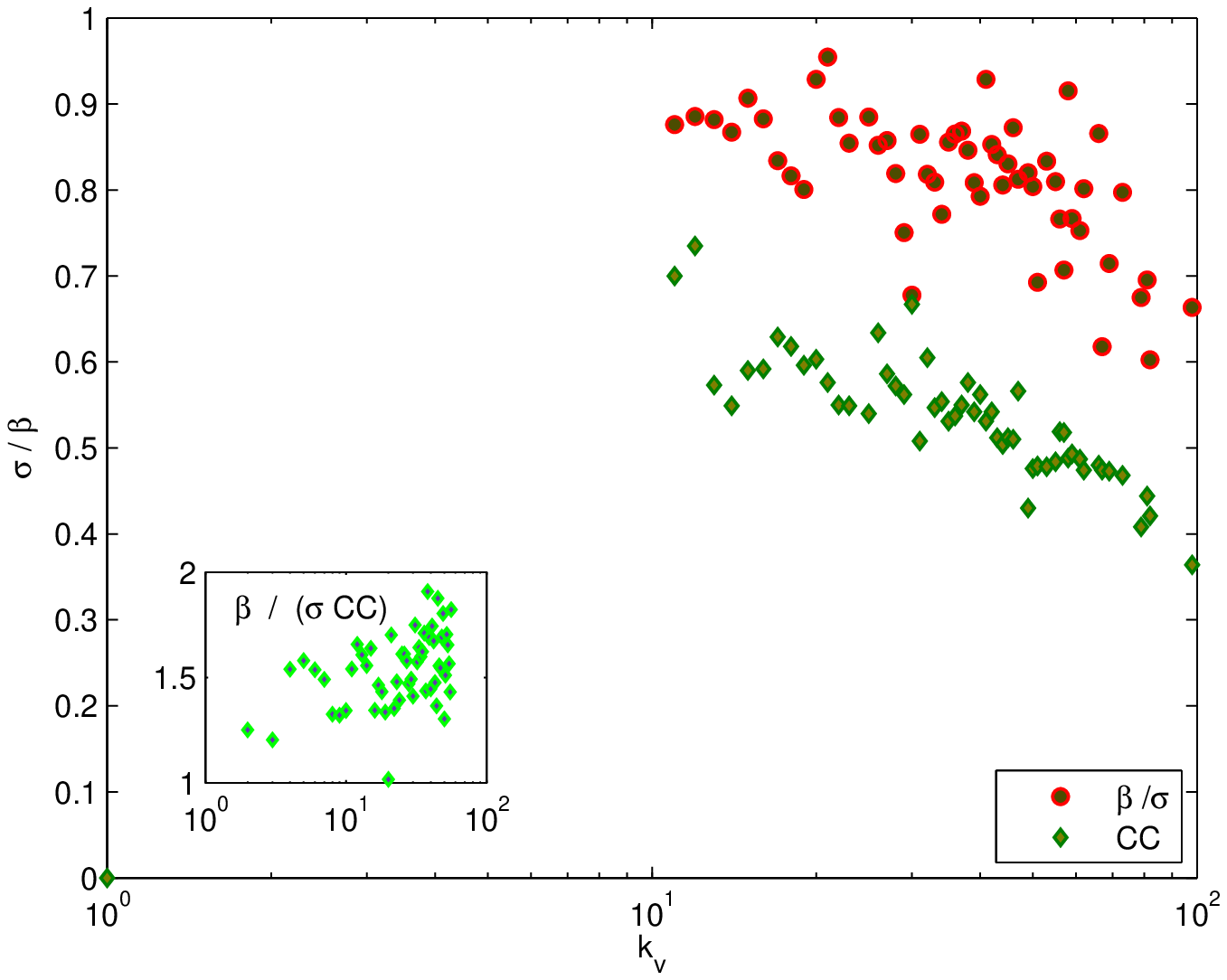}}
	\caption{(Color online)
		Rations 
		$\frac{\beta_{k}}{\sigma_{k}}$ and 
		$\frac{\beta_{k}}{\sigma_{k} \text{CC}_{k}}$ (inset)
		where $k$ is the degree of the victim.
	}
	\label{fig:bds-bdsCCXk}
	\end{center}
\end{figure}
%
%
%
Network type plays a key role on gossip spreading.
For example, co-authorship networks are formed by people 
coauthored in the same paper, and for this reason 
authors of a paper are all connected to each other.
Hence it is locally clique-like.
So gossips spread mostly through on small paths, 
i.e. more dense connectivity inside small groups.
This is a characteristic property of co-occurrence networks. 
This can be verified with the parallelism of the clustering coefficient and gossip spread rates in the undirected case, i.e. $\sigma / \text{CC}$.

However face-to-face proximity networks are generally formed by 
a person in the centre and other people know each other 
through the person in the center. 
The components of these networks are like cascades, that is one person passes to another, 
rather than fully connected cliques where 
one person have access to almost all. 
For these networks, although gossip spread rates are very high (around 0.9), clustering coefficient values are lower than ones in co-authorship networks.

\reftbl{tbl:coefficients} has the coefficients of the overall networks.
The co-occurrence group have CC in the range of $0.44-0.64$
where as that of social patterns group is much smaller and 
in a range of $0.36-0.56$.
The ratio of gossip spread to clustering coefficient 
$\sigma / \text{CC}$ is also an indicator.
For the co-occurrence group has smaller values then social patterns.
More interestingly, gossip propagation in the weighted model has 
more differential power. 
$\beta / \text{CC}$ values of the co-occurrence group are all below $1$. 
On the other hand those of the social pattern group are all above $1$.
Just a comparison, miscellaneous networks happen to have much higher 
$\sigma / \text{CC}$ and $\beta / \text{CC}$ values then the two groups.

%
\subsection{Variation by the Degree}

After investigating the clustering coefficient and 
gossip spread factors on the average, 
one questions the same figures as the degree changes.
\reffig{fig:bds-bdsCCXk} provides finer resolution to the degrees.
The rations of 
$\beta_{k} / \sigma_{k}$ and
$\beta_{k} / (\sigma_{k} \text{CC}_{k})$ are given as a function of degree $k$
where 
$\beta_{k}, \sigma_{k}, \text{CC}_{k}$ are 
the average clustering coefficient and 
spread factors 
of vertices of degree $k$
in weighted and corresponding unweighted networks, respectively.

The clustering coefficient $\text{CC}_{k}$ decreases as degree increases as in 
\reffig{fig:bds-bdsCCXk}.
$\beta / \sigma$ follows the pattern of first decrease to a minimum, then increase as in the case of $\sigma$ and $\beta$ in \reffig{fig:k0}.

The nodes with smaller degrees have relatively higher spread factors which 
shows that gossip about these nodes are spread to most of their neighbors. 
The  reason for this result is that 
these nodes are generally friends of highly connected nodes and 
their other small degree friends are also connected to 
the same highly connected nodes. 
This leads to high clustering coefficient for the node and 
yields higher spreading possibility as seen in the base model~\cite{Lind2007EPL}. 
This kind of connectivity is due to 
the power-law degree distribution frequently observed in real life networks
where many nodes have fewer connections while 
very few nodes have many 
connections~\cite{Barabasi1999Science}.

When weight is introduced, spread factors for low degree nodes are close to 
the values in unweighted base model. 
If a node with low degree is connected to a highly connected node, 
the weight of the edge is relatively important to the low degree node while 
it is generally less important for the highly connected node 
because of the fact that highly connected node's average weight is generally larger than the weight of the  connection with lower degree victim.  
This situation leads to the concept of being popular or important in the network. 
The nodes with small connectivity generally have low average strength, and 
when they are connected to a highly connected node 
which has greater average strength, 
the weight of the edge is not important to highly connected one 
as if highly connected nodes have few close friends. 
As a result of this situation, 
stop/spread decision generally turns to be to spread the gossips about ``weak'' nodes,
i.e. their gossip is spreaded by highly connected nodes since victim is not seen as an important friend by these nodes. 
This is an important feature of the network that has roots in social sciences as well.

\subsection{Strategies to Avoid Gossip}

Some network structures have superior properties in terms of gossip avoidance. 
Star-like network structure, where the victim is in the center is the best topology for gossip avoidance. 
Because of the structure, no friend of victim can communicate to each other without reaching the victim. 
For this structure, edge weights are not important; 
as the topology doesn't contain any triangles, there is no possibility of gossip.

At the other extreme, fully-connected graphs can be too gossipy. 
Because the topology fully enables gossip spread, 
only thing is the relative edge weights in the network. 
A change of an edge weight in the network may affect all the other nodes in the network in terms of gossip spreading. 
This is because of the fact that, decision function takes the relative importance of connections into considerations. 
So change of an edge value does not solely affect itself, 
but plays an important role on other edges due to relative evaluation of edge weights. 
Think about a scenario in real life; 
if you are closest friends of two person who know each other, 
you will not be gossiped by them. 
However if they get closer to each other and become closest friend of each other, 
then you will lose your position as the closest friend, 
although you did nothing wrong to destroy the connection with them. 
When this occurs, your connection is less important for them and 
they can gossip about you to each other.

According to our model, one can avoid gossip by applying some strategies: 

(i)~Eliminating triangles reduces the number of friends that can gossip about a victim. 

(ii)~Elimination of triangle cascades can also reduce the spread.
This can be done by having friends from different domains so that one from a domain does not know anybody from another domain.
With this a gossip can spread as far as on domain of friends go 
but cannot jump to another group.
As a general rule of thumb may be having islands of friend groups 
such that no intra-group communication is possible.
In real life we are in communities such as 
friends from high school, from college, from work.
As long as the communities do not overlap, 
spread of gossip is relatively under control.
The observations (i) and (ii) are valid for both weighted and unweighted networks.

(iii)~In weighted networks one can control the spread by means of the weights.
If victim $v$ have closer friendship to his friends $i, j$ 
than friendship of $i$ and $j$,
i.e. $w_{iv} > w_{ij}$ and $w_{jv} > w_{ji}$.
In a directed weighted network this has an interesting consequence:
What your friends think of you is important 
rather then what you think of them, i.e. $w_{iv}$ vs $w_{vi}$.

\subsection{Generated Weighted Networks}
\label{sec:GeneratedWeightedNetworks}

\begin{figure}
	\begin{center}
	\subfigure[GER]{
		\includegraphics[scale=0.4]{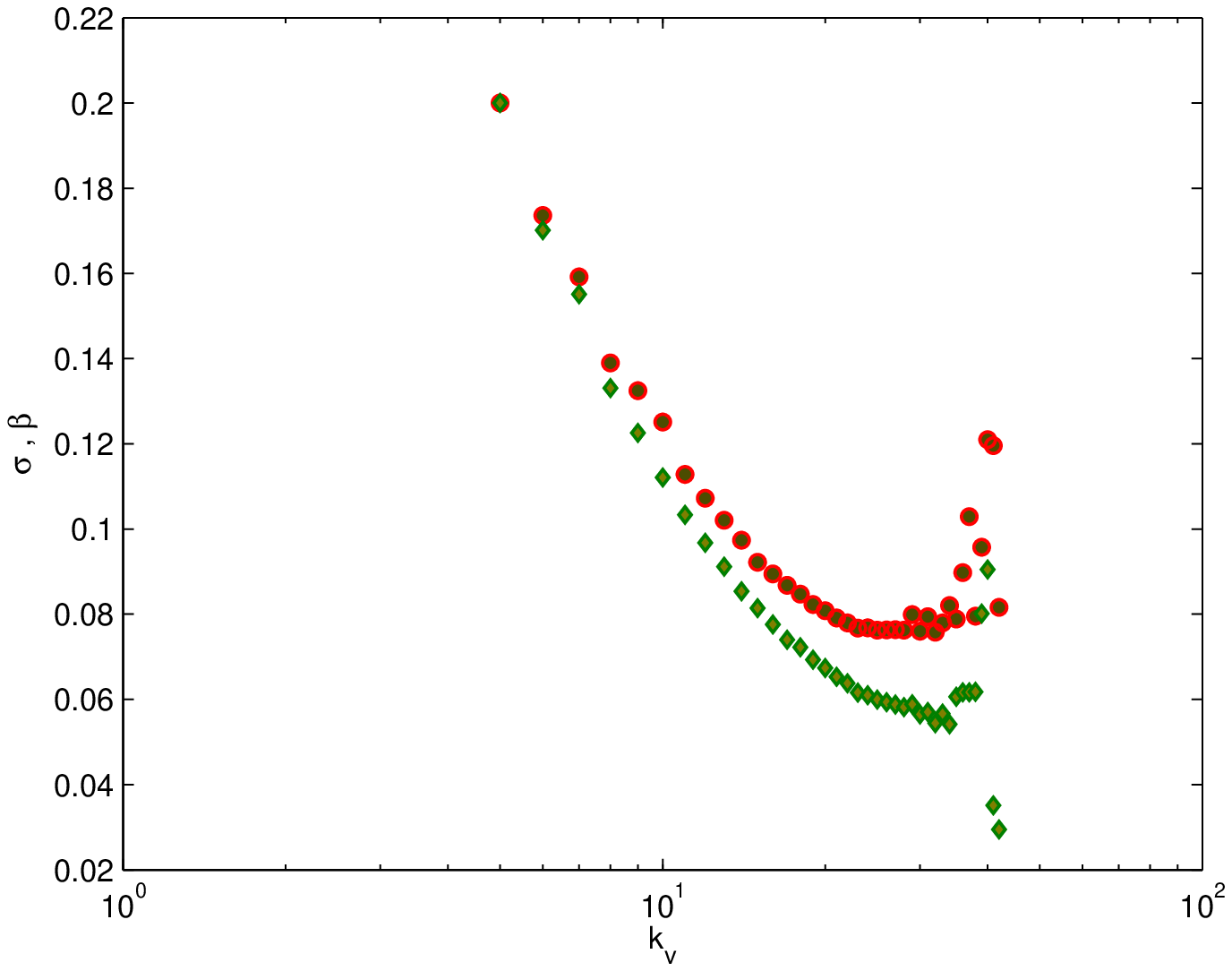}}
	\subfigure[GBA]{
		\includegraphics[scale=0.4]{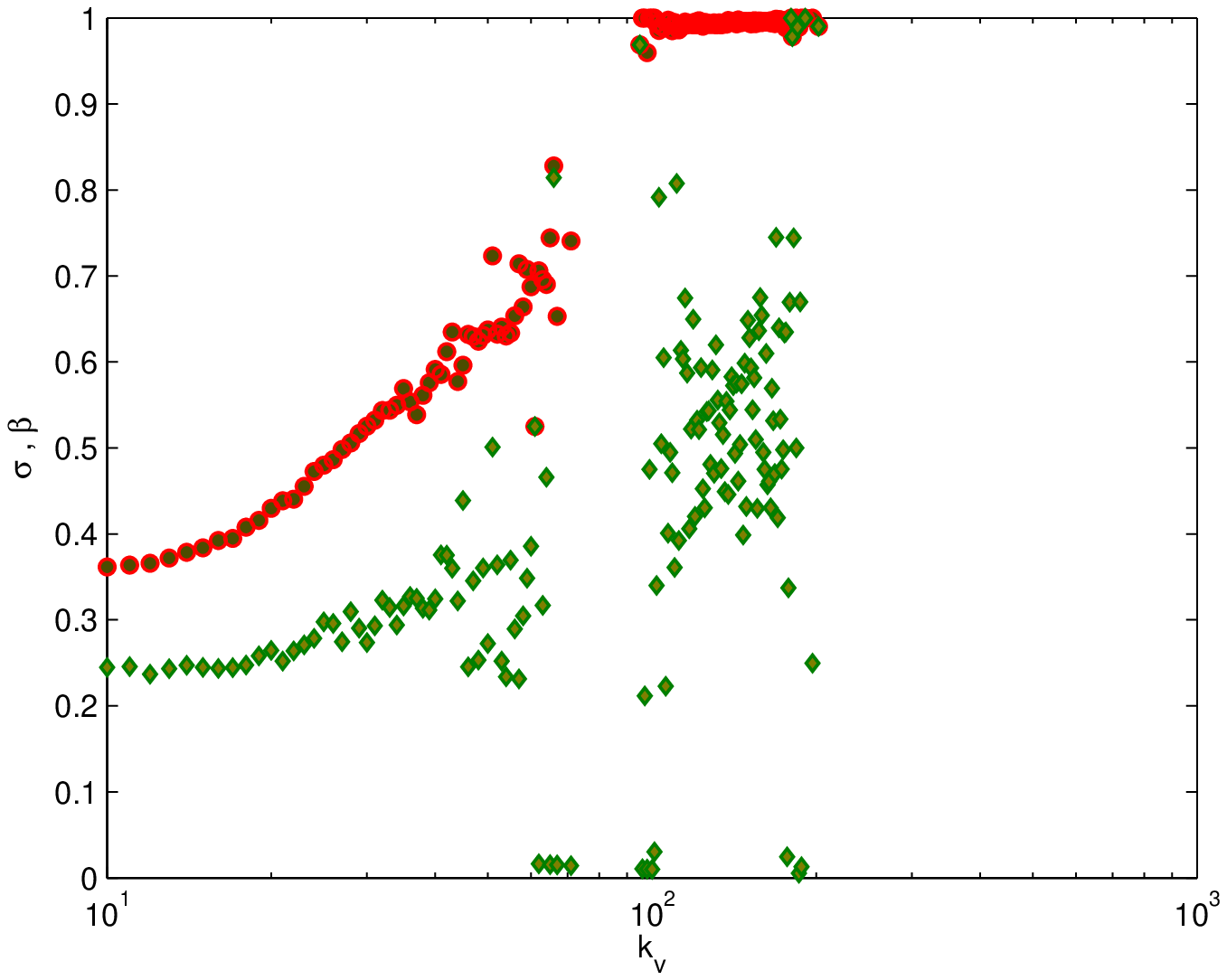}}
	\subfigure[GWS]{
		\includegraphics[scale=0.4]{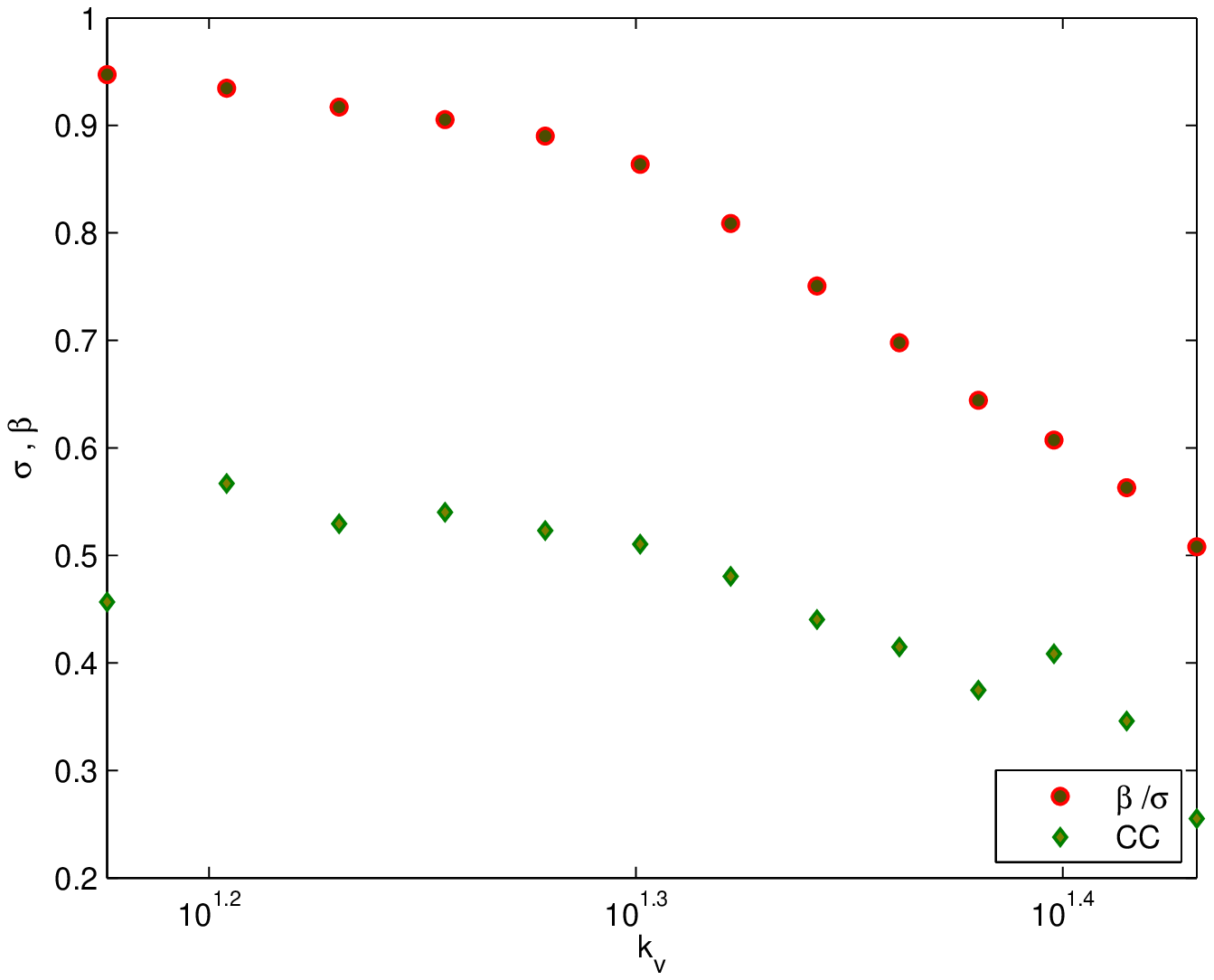}}		
	\caption{
		$\sigma_{k_{v}}$ and $\beta_{k_{v}}$ values for generated networks.
		The values are the averages of 50 realizations.
	}
	\label{fig:s-bXk_G-all}
	\end{center}
\end{figure}
So far weighted networks emprically obtained from real life are investigated.
One wonders gossip spread on the networks generated by means of well-known models such as 
Erdos-Renyi (ER), 
Barabasi-Albert (BA), and
Watts-Strogatz (WS)~\cite{Erdos1959RandomGraphs,Barabasi1999Science, Watts1998Nature}.

These networks are unweighted networks. 
In order to have edge weights, 
we use a Gaussian distribution, with mean 1 and variance 1,
to generate weight $w_{i}$ for node $i$.
Then edge weight $w_{ij}$ is assigned as the average of the node weights, 
i.e. $w_{ij} = 0.5 (w_{i} + w_{j})$.

All generated networks have $N = 200$ nodes and 
their generation parameters are set such that 
they the number of edges is around $M = 800$. 
These values of $N$ and $M$ are selected arbitrary.
For each model 50 networks are generated. 
The results of the average of these 50 realizations are reported in 
\reftbl{tbl:coefficients} and 
\reffig{fig:s-bXk_G-all}.

The small-world networks of Watts-Strogatz (WS), 
denoted as GWS, 
are investigated 
first~\cite{Watts1998Nature}.
$N = 200$, $k = 4$, and rewiring probability $p = 0.1$ are used. 
These networks are regular networks with some re-wired links. 
As these networks have very few triangles,  
their gossip spread factor is not high. 
Even we increased the number of re-wiring, 
overall gossip spread is not large due to topology. 

Power-law degree distribution networks generated by Barabasi-Albert (BA) model
are denoted by GBA~\cite{Barabasi1999Science}.
Network size is $N=200$ and  
initial clique size $m_{0} = 10$. 
Each newly added node is connected to 4 existing nodes in the network (i.e. each node increase the number of edges by $4$).
We see high values of gossip spreads. 
This is due to the topology of BA network, 
where the core of the network is a fully connected graph and 
newly inserted nodes have high preferential attachment,
i.e. tend to connect to highly connected nodes. 
Network generation process creates highly connected hubs, 
each having many connections who know each other. 
Although weights play an important role in decision of gossip spreading, 
topology adds more to the gossip spreading here and 
overall gossip spread is greater. 

Finally, random networks of Erdos-Renyi, 
denoted by GER,
are investigated~\cite{Erdos1959RandomGraphs}.
$N = 200$ with connection probability $p = 0.04$ are used.
As random networks do not have scale-free property, they do not have high clustering coefficient and preferential attachment.
For this reason, topology of random networks is not suitable for gossip spreading; i.e. there are very number of triangles in these networks. 
Gossip spread rates for both unweighted and weighted models are very close to each other. 
This paralellism is the result of both randomness of weight distribution and the lack of triangles in these networks.

%
%

\section{Conclusion}
In this paper, we propose a gossip spread model in weighted networks. 
Weight effect reveals more information about the connection patterns, 
degree distributions and topology of the network. 
As a result of power-law degree distribution, 
both high-degree and low-degree victims are more vulnerable to being gossiped 
even when the weight promotes not spreading the gossip. 
The first reason is different evaluation of an edge by two parties; 
one node may see the connection as ``important'' 
while the other one may see as ``not so important''. 
Low degree nodes have highly connected neighbors for whom 
the lower degree nodes are not seen as ``important''. 
The second reason for this behavior is a threshold degree after which 
further connections are not closer friends to victim. 
These connections can be seen as the result of 
preferential attachment in real networks and 
most of the time they are not ``close friends'' of the victim but 
rather the connections those are made due to high degree of victim. 
We can interpret this threshold degree as the optimal friendship capacity 
one can manage.

An interesting finding is another threshold degree value, 
after which a node is accepted as an ``important'' node in the network. 
This degree threshold is at the point where 
cross-over occurs in spread factor graph,
i.e. when the graph of  decision according to spreader's average weight 
cross-overs the graph of decision according to victim's average weight.

\begin{acknowledgments}
	This work was partially supported 
	by Bogazici University Research Fund, BAP-2008-08A105, 
	by the Turkish State Planning Organization (DPT) TAM Project, 2007K120610,
	and
	by COST action MP0801.
\end{acknowledgments}


\bibliography{MurselGossip}{}

\end{document}